\documentclass{article}
\usepackage{cite}
\usepackage{amsmath,amssymb,amsfonts}
\usepackage{algorithmic}
\usepackage{graphicx}
\usepackage{textcomp}
\usepackage{xcolor}
\usepackage[caption=false,font=footnotesize]{subfig}
\usepackage[export]{adjustbox}
\usepackage{circuitikz}
\usepackage{array}
\usepackage{multirow}
\usepackage{balance}
\usepackage{breqn}
\usepackage{steinmetz}
\usepackage{authblk}

\usepackage[margin=1in]{geometry}

\definecolor{myblue}{rgb}{0, 0.4470, 0.7410}
\definecolor{mygray}{rgb}{0.7, 0.7, 0.7}
\tikzset{>=latex} 

\title{Practical Superdirective and Efficient Rectenna\\for Low-Power RF Energy Harvesting}

\author[1]{Aaron M. Graham}
\author[1]{Stylianos D. Asimonis}
\affil[1]{Centre for Wireless Innovation (CWI), Queen's University Belfast, BT3
	9DT Belfast, U.K.}
\affil[ ]{\textit {\{agraham63, s.asimonis\}@qub.ac.uk}}

\date{}

\begin{document}
	
	\maketitle
	
	\newcommand{\ww}{2}
	\newcommand{\hh}{2.5}
	
	\begin{abstract}
		This paper presents the design and analysis of a superdirective rectenna optimized for enhanced RF-to-DC efficiency and realised gain, featuring direct impedance matching between the antenna and the rectifier. Employing passively loaded, low profile strip dipoles on a low-loss substrate, the rectenna achieves a realised gain of $6.9$ dBi and an RF-to-DC efficiency of $22.1\%$ at $-20$ dBm.
	\end{abstract}
	
	\textbf{Keywords:}
	Energy harvesting, Rectennas, Directive antennas, Superdirective antenna arrays
	
	\section{Introduction}
	Superdirective antennas are pivotal components in modern RF and microwave engineering \cite{Lynch2024, 10238734}, particularly in applications involving wireless power transmission and RF energy harvesting. Rectennas, devices that combine an antenna and a rectifying circuit (rectifier) in a single unit, convert RF energy into DC power \cite{7433469,Eid2021}. The RF-to-DC efficiency of rectennas is highly dependent on the matching between the antenna's input impedance and the rectifier's impedance, which is typically highly capacitive or inductive, as well as the diode's characteristics to handle low-power signals effectively. Superdirective antennas, which exhibit a highly directional radiation pattern achieved by closely spaced elements with controlled amplitude and phase excitation, often suffer from low radiation efficiency and typically present highly inductive or capacitive input impedance, making them impractical. However, recent studies show that by carefully adjusting antenna size and using reactance loads, the design of impedance matched and highly radiation-efficient superdirective antennas is possible \cite{Moore2024Low,AssimonisHow2023}. 
	
	Superdirective rectennas integrate the principles of superdirective antennas with rectenna technology, aiming to enhance the directivity and sensitivity of RF energy harvesting systems \cite{9686565}. Additionally, since both the rectifiers and superdirective antennas present highly inductive and/or capacitive input impedance, the design of a superdirective rectenna where the superdirective antenna is directly connected to the rectifier is possible, reducing extra losses by eliminating the need for an impedance matching network.
	
	The contribution of this work is the design of a practical superdirective rectenna with an electrical size of $1.6$, combining the concept of passively loaded strip dipoles printed on a low-loss substrate. The proposed rectenna was analyzed in terms of RF-to-DC efficiency ($22.1\%$ at $-20$ dBm or for $0.35 \, \mu\text{W/cm}^2$), directivity ($7.3$ dBi with a maximum of $7.6$ dBi, based on Harrington's limit), gain ($7.1$ dBi), and realized gain ($6.9$ dBi). The antenna of this rectenna was designed to be directly impedance-matched to the rectifier.

	\section{Rectifier}
	
\begin{figure*}[!t!]
		\centering
		\subfloat[]{
			\begin{adjustbox}{valign=c} 
				\centering
				\begin{circuitikz}[american, line width=0.5pt, scale=0.7, transform shape]
					\ctikzset{bipoles/thickness=1}
					\draw
					(0,0) 
					++(0,\hh) to[sV, l_=\large$P_{in}$] ++(0,-\hh) ++(0,\hh) 
					to[R, l=\large$Z_{a}$, -o] ++(\ww,0)
					to[D, l=SMS7630-005L] ++(\ww,0)
					to[C, l=\large$C$] ++(0,-\hh) 
					++(0,\hh) -- ++(\ww,0)
					to[R, l=\large$R$] ++(0,-\hh) -- (0+\ww,0) to[short, o-] (0,0)
					(\ww-0.25,-0.5) node[below] {\large$Z_{r}$} -- ++(0,1) [-latex] -- ++(0.5,0); 
					;
				\end{circuitikz}
			\end{adjustbox}
			\vphantom{
				\includegraphics[width=0.3\linewidth,valign=m]{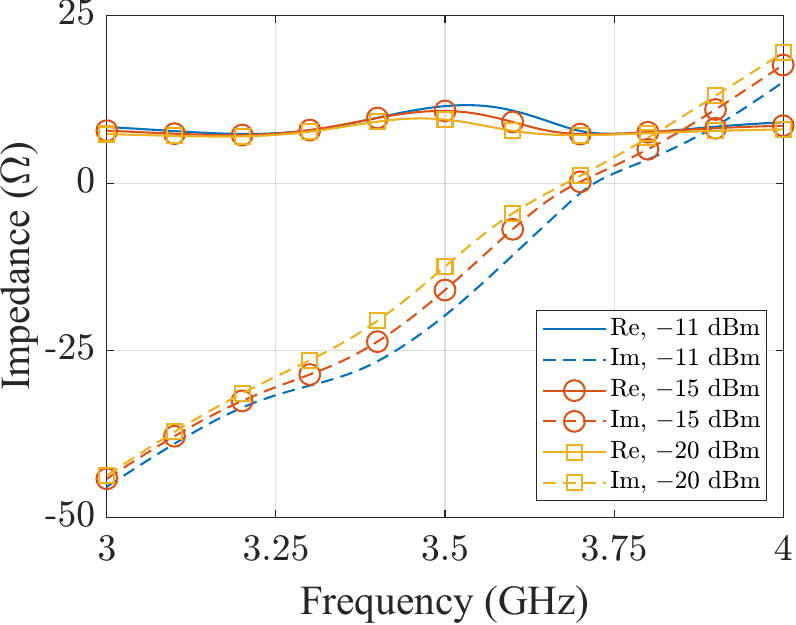}
			}
			\label{fig1a}
		}
		\hfil
		\subfloat[]{
			\includegraphics[width=0.3\linewidth,valign=m]{Figs/Fig02_Single_Impedance}
			\label{fig1b}
		}
		\hfil
		\subfloat[]{
			\includegraphics[width=0.3\linewidth,valign=m]{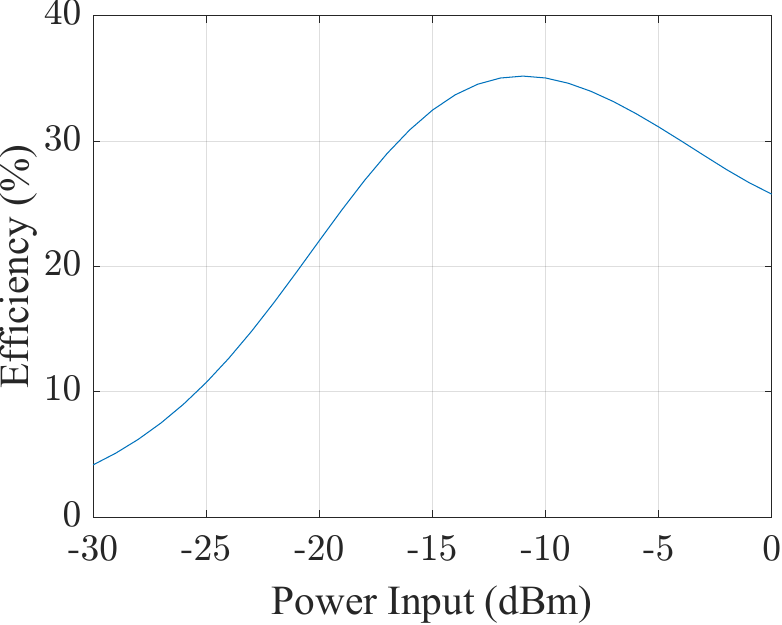}
			\vphantom{
				\includegraphics[width=0.3\linewidth,valign=m]{Figs/Fig02_Single_Impedance}
			}
			\label{fig1c}
		}
		\\
		\subfloat[]{
			\includegraphics[width=0.3\linewidth,valign=m]{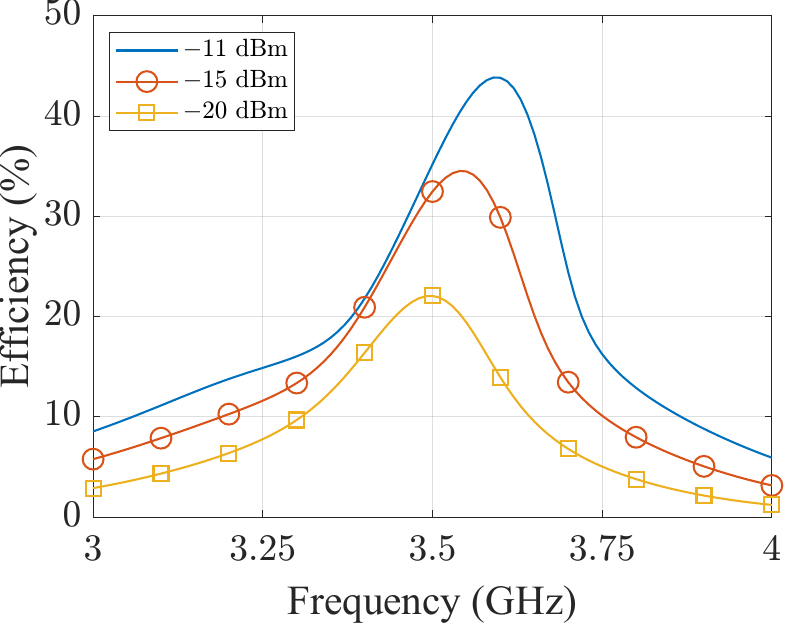}
			\label{fig1d}
		}
		\hfil
		\subfloat[]{
			\includegraphics[width=0.3\linewidth,valign=m]{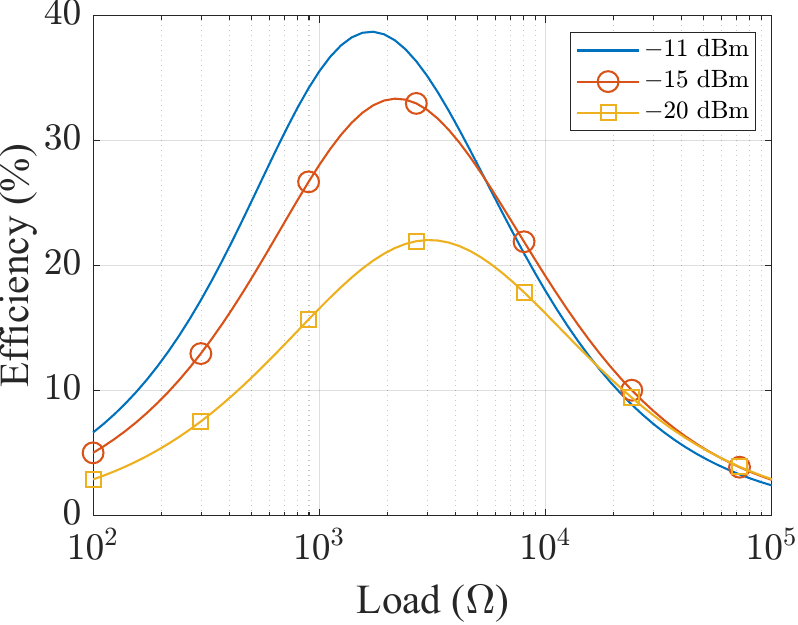}
			\label{fig1e}
		}
		\hfil
		\subfloat[]{
			\includegraphics[width=0.3\linewidth,valign=m]{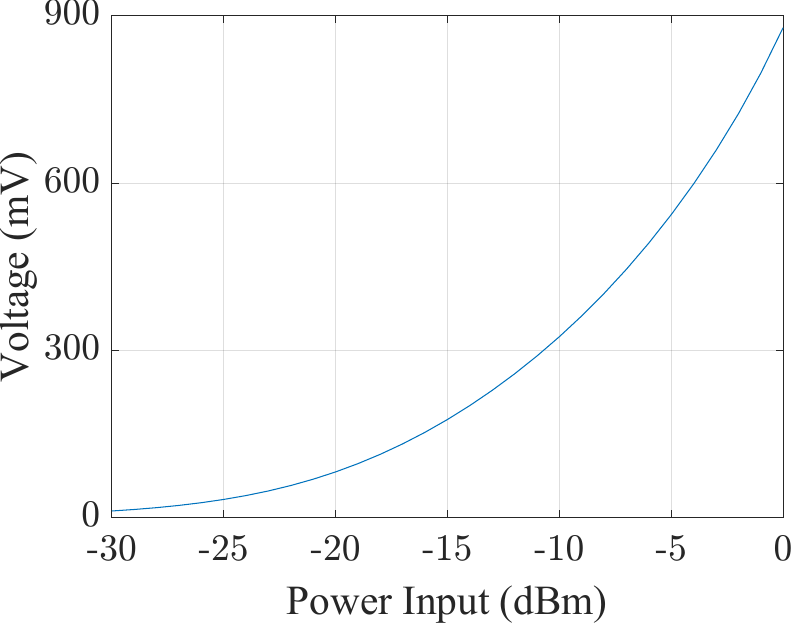}
			\label{fig1f}
		}
		\caption{ 
			Comprehensive analysis of an RF-to-DC rectifier: (a) Circuit design and impedance estimation, (b) Simulated input impedance versus frequency for various power input levels, © RF-to-DC efficiency versus power input at 3.5 GHz, (d) RF-to-DC efficiency versus frequency for various power input levels, (e) RF-to-DC efficiency versus load for various power input levels, and (f) Output DC voltage variation with power input. 
		}
		\label{fig1}
	\end{figure*}

	A rectifier, designed to operate at $3.5$ GHz with a low-power input of $-20$ dBm, features a coplanar geometry. This design eliminates the need for a ground plane and simplifies the connection with the strip dipole antenna. The substrate of choice is Rogers RO4350B, which is characterized by a relative permittivity $\epsilon_r = 3.48$, a loss tangent $\tan \delta = 0.0037$, and a thickness of $0.51$ mm. 
	As depicted in Fig. \ref{fig1a}, the rectifier consists of a series circuit for half-wave RF-to-DC rectification. This circuit incorporates a single Schottky diode (SMS7630-005L), a capacitor, and a load. The design omits a matching network, as direct impedance matching of the antenna to the rectifier will be addressed in subsequent steps. 
	The rectifier's input impedance, which varies versus power input, operating frequency, and output load, has been estimated via harmonic balance analysis using ADS software from Keysight Technologies. The estimated impedance is $Z_r = 10.1-j12.9$ $\Omega$. 
	The simulation aimed to determine the rectifier's input impedance and output load under conditions that maximize RF-to-DC efficiency. The optimal conditions were found with an output load of $3000$ $\Omega$, which resulted in a maximum RF-to-DC efficiency of $22.1\%$.

	Fig. \ref{fig1b} illustrates the simulated input impedance in terms of real (Re) and imaginary (Im) parts versus frequency for various power input levels. It is evident that the rectifier exhibits capacitive and non-linear characteristics. At $ 3.5 $ GHz, the Re and Im parts of the input impedance vary from $ 9.5 $ $\Omega$ to 10.5 $\Omega$ and from $ -19.8 $ $\Omega$ to $ -12.5 $ $\Omega$, respectively, for power input levels ranging from $-20$ dBm to $-11$ dBm.
	Subsequently, the input impedance of the generator was fixed at $Z_a = 10.1 + j12.9$ $\Omega$ to conjugate match with the rectifier. The RF-to-DC efficiency versus power input at $ 3.5 $ GHz was then estimated. Fig. \ref{fig1c} presents the simulated results: the efficiency increases from $ 4.2\% $ to $ 35.2\% $ for power input levels ranging from $ -30 $ dBm to $ 0 $ dBm. Notably, the efficiency is $ 22.1\% $ for a $ -20 $ dBm power input.
	
	Fig. \ref{fig1d} and \ref{fig1e} illustrate the RF-to-DC efficiency versus frequency (with a fixed load of $ 3000 $ $\Omega$) and load (at $ 3.5 $ GHz), respectively, for various power input levels. These figures are generated when the rectifier is powered by an impedance-matched source, i.e., $Z_a = 10.1 + j12.9$ $\Omega$. As expected, a higher power input results in higher efficiency. However, this increase in efficiency is accompanied by a slight increase in frequency (as shown in Fig. \ref{fig1b}) and a lower load (as depicted in Fig. \ref{fig1e}). The output DC voltage, depicted in Fig. \ref{fig1f}, varies from $ 10 $ mV to $ 880 $ mV as the power input increases from $ -30 $ dBm to $ 0 $ dBm. Specifically, for a power input of $ -20 $ dBm, the output DC voltage is approximately $ 81 $ mV.

	\section{Superdirective Rectenna}
	
	The low-profile superdirective antenna, designed for the 5G sub-6 GHz frequency band as discussed in \cite{Moore2024Low,AssimonisHow2023}, is to be adapted for the design of a superdirective rectenna. This design includes a two-element array of strip dipoles with varying lengths $L_1$ and $L_2$, and widths $W_1$ and $W_2$, positioned side by side at a distance $d$. To achieve superdirectivity, the first dipole element is directly excited by a sinusoidal voltage signal, while the second element is unexcited but incorporates a purely reactive load $Z_L$ connected at the mid-point of the strip between its arms. Consequently, this second element functions as a parasitic loaded element, significantly influencing the array's radiation pattern and directivity.
	
\begin{figure*}[!t!]
		\centering
		\subfloat[]{
			\begin{adjustbox}{valign=c} 
				\includegraphics[width=0.3\linewidth,valign=m]{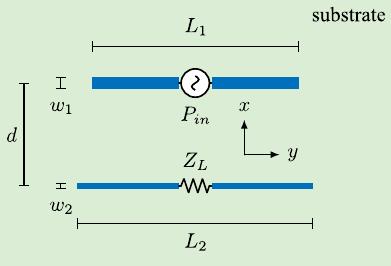}
				\label{fig2a}
			\end{adjustbox}
			\vphantom{
				\includegraphics[width=0.3\linewidth,valign=m]{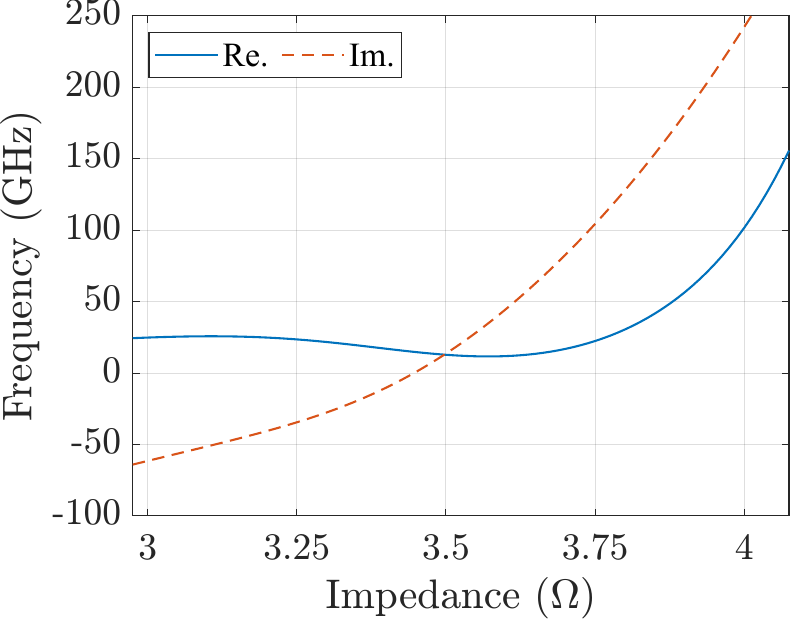}
			}
		}
		\hfil
		\subfloat[]{
			\includegraphics[width=0.3\linewidth,valign=m]{Figs/antenna_Imp_001}
			\label{fig2b}
		}
		\hfil
		\subfloat[]{
			\includegraphics[width=0.3\linewidth,valign=m]{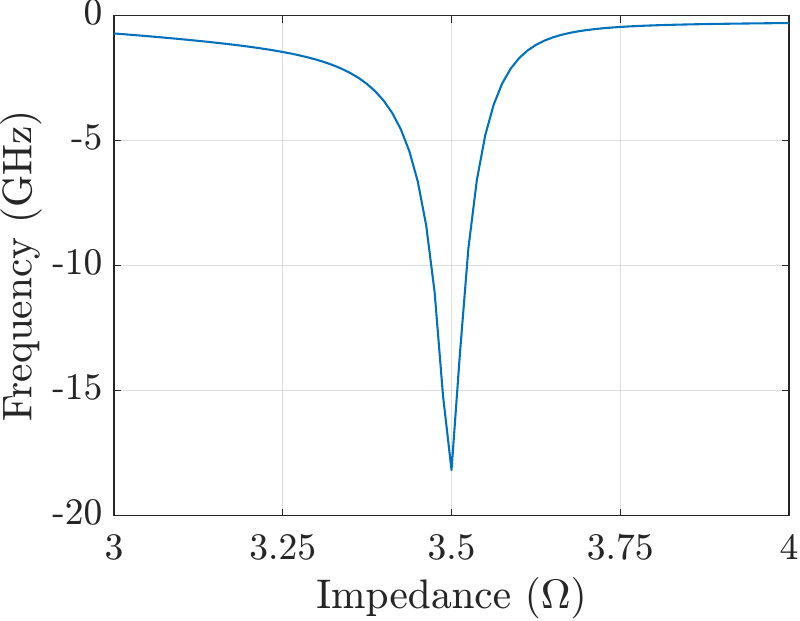}
			\vphantom{
				\includegraphics[width=0.3\linewidth,valign=m]{Figs/Fig02_Single_Impedance}
			}
			\label{fig2c}
		}
		\\
		\subfloat[]{
			\includegraphics[width=0.3\linewidth,valign=m]{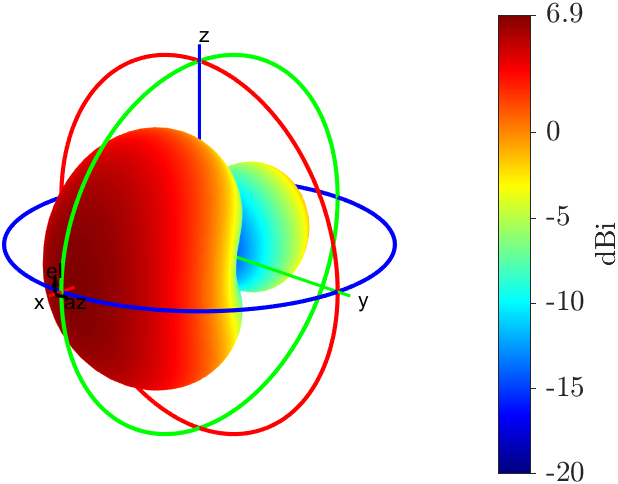}
			\label{fig2d}
		}
		\hfil
		\subfloat[]{
			\includegraphics[width=0.3\linewidth,valign=m]{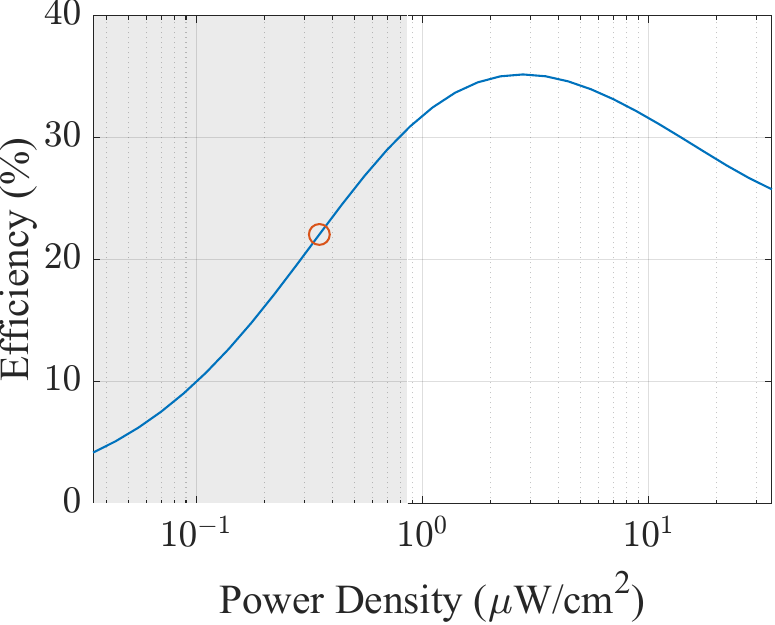}
			\label{fig2e}
		}
		\hfil
		\subfloat[]{
			\includegraphics[width=0.3\linewidth,valign=m]{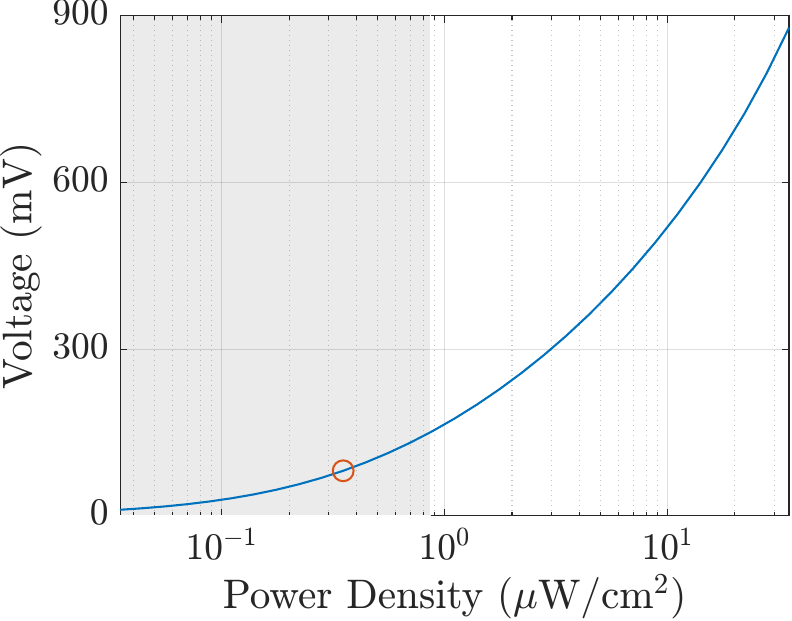}
			\label{fig2f}
		}
		\caption{ 
			Figure 2: Performance characteristics of the designed superdirective rectenna system. (a) antenna array design, (b) complex antenna impedance, (c) reflection coefficient versus frequency, (d) antenna realized gain, (e) RF-to-DC efficiency versus power density, and (f) DC output voltage versus power density.	 
		}
		\label{fig2}
	\end{figure*}

	The antenna array (Fig. \ref{fig2a}) was designed with the aim of achieving superdirectivity at a frequency of $3.5$ GHz. The design incorporated strip dipoles, which were positioned side by side on a Rogers RO4350B substrate. This substrate had a relative permittivity ($\epsilon_r$) of $3.48$, a loss tangent ($\tan \delta$) of $0.0037$, and a thickness of $0.51$ mm. The metallic components of the dipole strips were modeled of copper, which had a conductivity of $5.8 \times 10^7$ S/m and a thickness of $35$ µm. The interelement distance ($d$), measured from the middle of one strip to the middle of the adjacent strip, was set to $0.0815\lambda$. In the operational setup, the first strip dipole was excited while the second was loaded with a load ($Z_L$). The load and the dimensions of the dipoles were determined through an optimization process. The simulation of the antenna was carried out using the Antenna Toolbox of MATLAB, with the Method of Moments applied for accurate results.
	
	Mathematically, the optimization problem is expressed as:
	
	\begin{equation}
		\begin{aligned}
			\underset{
				\left\lbrace L_1,L_2,W_1,W_2,Z_L \right\rbrace 
			}{\text{Maximize}}
			& \quad f \left(L_1,L_2,W_1,W_2,Z_L\right) \\
			\text{subj. to:} 
			& \quad L_1, L_2 \in \left[0.3\lambda, 0.7\lambda\right],\\
			& \quad W_1, W_2 \in \left[10^{-3}\lambda, 10^{-1}\lambda\right],\\
			& \quad Z_L \in \left[-j 10~\mathrm{k}\Omega, j +10~\mathrm{k}\Omega\right],
		\end{aligned}
	\end{equation}
	where $f$ represents the realized gain function, with reference impedance $Z_r = 10.1 - j12.9 \, \Omega$, which is the the rectifier's input impedance, hence,
	\begin{equation}
		f = (1 - \left| \Gamma \right|^2 ) G,
	\end{equation}	
	where $G$ is the antenna gain and
	\begin{equation}
		\Gamma = \frac{Z_a - Z_r^{*}}{Z_a + Z_r},
	\end{equation}	
	where $\left\lbrace *\right\rbrace $ denotes the conjugate matched.
	The optimization results are as follows: 
	$L_1 = 0.4392\lambda$, 
	$L_2 = 0.5097\lambda$,
	$w_1 = 0.0189\lambda$,
	$w_2 = 0.0192\lambda$, and
	$Z_L = -252.46~\Omega$.
	The reactance of $-252.46 \, \Omega$ corresponds to a capacitance of $
	C = \left(2\pi \times 3 \times 10^9 \times 252.46\right)^{-1} = 0.18 \, \text{pF}. $

	The complex antenna impedance is estimated and presented in Fig.~\ref{fig2b}. At 3.5 GHz, it is 
	$
	Z_a = 12.8 + j13.67 \, \Omega.
	$
	The antenna is connected to the rectifier, and the corresponding reflection coefficient is 
	$
	\Gamma = -18.2 \, \text{dB}.
	$
	Additionally, the reflection coefficient versus frequency is depicted in Fig.~\ref{fig2c}. Therefore, with a $-20 \, \text{dBm}$ power input, the antenna is impedance matched to the rectifier from $ 3.47 $ GHz to $ 3.52 $ GHz, resulting in a fractional bandwidth of $ 1.43\% $.

	The antenna realized gain $ G $ is depicted in Fig.~\ref{fig2d}, and the maximum of $6.9$ dBi occurs towards the $x$ direction. Therefore, the loaded parasitic element acts as a reflector.
	Additionally, the corresponding directivity and gain are $7.3$ dB and $7.1$ dBi, respectively.
	Furthermore, according to Harrington's limit \cite{Harrington1958}, the maximum directivity, $D_{\text{max}}$, of a lossless antenna that completely fills a sphere with radius $R$, is given by the equation:
	\begin{equation}
		D_{\text{max}} = (kR)^2 + 2kR,
	\end{equation}
	where $k = \frac{2\pi}{\lambda}$ is the wave number. In this case, the electrical size of the antenna is $kR = 1.6$, therefore, $D_{\text{max}} = 7.6$ dBi.
	Hence, this rectenna achieves maximum directivity very close to the maximum theoretical directivity assuming an ohmic-less antenna.
	Most importantly, this rectenna presents high gain ($7.1$ dBi) and is impedance matched directly to the rectifier, resulting in high realized gain.

	Next, the RF-to-DC efficiency of the rectenna was estimated. For this, the power density $S$ was calculated based on the realized gain of the antenna, i.e.,
	\begin{equation}
		S = \frac{P_{\text{in}}}{A_{\text{e}}},
	\end{equation}
	where,
	\begin{equation}
		A_{\text{e}} = \frac{4\pi}{\lambda} \, G
	\end{equation}
	is the rectenna’s effective area \cite{balanis2016antenna}. The results are depicted in Fig.~\ref{fig2e}. Based on long-term RF electromagnetic field (EMF) measurements in the European Union \cite{Gajsek2015}, the power density ranges from $0.0017$ to $0.8594 \, \mu\text{W/cm}^2$, and the grey area denotes this region. Hence, the superdirective is capable of operating inside this region with a maximum efficiency of $31\%$ at $0.8594 \, \mu\text{W/cm}^2$. For $0.35 \, \mu\text{W/cm}^2$ (red circle in Fig.~\ref{fig2e}), the rectenna can harvest $-20 \, \text{dBm}$ and presents an efficiency of $22.1\%$, as mentioned.
	Similarly, the DC output voltage versus power density is depicted in Fig.~\ref{fig2f}. Within the $0.0017$--$0.8594 \, \mu\text{W/cm}^2$ region, the superdirective rectenna is capable of delivering up to $153$ mV. At $0.35 \, \mu\text{W/cm}^2$ (marked by the red circle in Fig.~\ref{fig2f}), the output voltage is $81$ mV.

	
	\bibliographystyle{IEEEtran}
	\bibliography{mybib}
	
	\vspace{12pt}
	
\end{document}